\documentclass[aps,11pt]{revtex4}

\def\[{\left\lbrack}
\def\]{\right\rbrack}

\def\({\left(}
\def\){\right)}

\newcommand{\be}{\begin{equation}}
\newcommand{\ee}{\end{equation}}
\newcommand{\ea}{\end{eqnarray}}
\newcommand{\ba}{\begin{eqnarray}}

\begin{document}

\title{Eliminating the chiral anomaly {\it via} symplectic embedding approach}

\author{A. C. R. Mendes$^a$\footnote{\sf E-mail: albert@ufv.br}, C. Neves$^b$\footnote{\sf E-mail: cneves@fisica.ufjf.br}, W. Oliveira$^b$\footnote{\sf E-mail: wilson@fisica.ufjf.br}} 
\affiliation{${}^{a}$Universidade Federal de Vi\c cosa, Campus Rio Parana\'{\i}ba, 38810-000, Rio Parana\'{\i}ba, MG \\
${}^{b}$Departamento de F\'{\i}sica,ICE, Universidade Federal de Juiz de Fora,\\
36036-330, Juiz de Fora, MG, Brazil\\
\bigskip
\today}

\begin{abstract}
The quantization of the chiral Schwinger model $(\chi QED_{2})$
with one-parameter class Faddeevian regularization is hampered by the chiral anomaly,
{\it i.e.}, the Gauss law commutator exhibits Faddeev's anomaly. To
overcome this kind of problem, we propose to eliminate this
anomaly by embedding the theory through a new gauge-invariant
formalism based on the enlargement of the phase space with the 
introduction of Wess-Zumino(WZ) fields and the symplectic approach\cite{ANO,Nucl}. This process
opens up a possibility to formulate different, but dynamically
equivalent, gauge invariant versions for the model and also gives a
geometrical interpretation to the arbitrariness presents on the BFFT 
and iterative conversion methods. Further, we observe that the elimination
of the chiral anomaly imposes a condition on the chiral parameters present 
on the original model and on the WZ sector.

\end{abstract}

\maketitle

\newpage

\section{Introduction}

It has been shown over the last decade that anomalous gauge
theories in two dimensions can be consistently and unitarily
quantized for both
Abelian\cite{JR,RR,many} and non-Abelian\cite{RR1,LR} cases. In this scenario, the two dimensional model that has been extensively studied
is the chiral Schwinger model(CSM)\cite{JR}. In these early works were considered to study an effective action with a simple one-parameter
class of regularization. The consequences of these constraint structures are that the $a>1$ class presents, besides the massless
excitation also a massive scalar excitation
($m^2=\frac{e^2a^2}{a-1}$) that is not found on the $a=1$ class.
In order to elucidate the physical spectrum, this model was
analyzed by a variety of methods\cite{RR,GRR,FK,MS,IJMP}. In
Refs.\cite{FK,MS,IJMP} a gauge invariant formulation of the model
with Wess-Zumino(WZ) fields was studied, as suggested in
Ref.\cite{FS}.

Despite this spate of interest, Mitra \cite{PM} proposed a new and
surprising bosonized action for the Schwinger model with a new
regularization prescription, that is different from those involved
in the class of models studied earlier. In this paper he proposed
a new (Faddeevian) regularization class, incarnated by a
conveniently mass-like term which leads to a canonical description
with three second class constraints, in the Dirac's context. In
this paper, it has been shown that the Gauss law commutator
exhibits Faddeev's anomaly. This model has an advantage because
the Faddeevian mechanism\cite{faddeev}, which is related to the
anomalous Gauss law algebra in the anomalous gauge theories, works
well. Recall that in \cite{JR} and \cite{RR}, the Hamiltonian
framework was structured in terms of two classes with two and four
second class constraints respectively. Mitra's work brings a clear
interpretation about the reason leading the bosonization ambiguity
to fit into 3 distinct classes, classified according to the number
of constraints present in the model. Recently, an extension of the
Ref.\cite{PM} to an one-parameter class of solutions was proposed
by one of us in \cite{AINW} in order to study the restrictions
posed by the soldering formalism\cite{CW} over this new
regularization class. In this context, the gauge field becomes
massive once again and its dependence on the ambiguity parameter
was shown to be identical to that in Ref.\cite{JR}, while the
massless sector however is more constrained than its counterpart
in \cite{JR}, which corroborates the Mitra's finding
outs\cite{PM}.

In the present paper, the chiral Schwinger model ($\chi QED_{2}$) with
one-parameter class Faddeevian regularization will be reformulated
as a gauge invariant theory in order to eliminate the chiral
anomaly that obstructed the quantization procedure. It is done
just enlarging the phase space with the introduction of WZ fields.
To this end, we chose to use a new gauge-invariant formalism, that
is an extension of the symplectic gauge-invariant formalism\cite{ANO,Nucl}. This formalism is
developed in a way to handle constrained systems, called the
symplectic framework \cite{JF,JW}. The basic object behind this
formalism is the symplectic matrix: if this matrix is singular,
the model presents a symmetry. This is achieved introducing an
arbitrary function $(G)$, written in terms of the original and WZ
variables, into the zeroth-iterative first-order Lagrangian.

In section II, we introduce the symplectic embedding formalism in
order to settle the notation and familiarize the  reader with the
fundamentals of the formalism. In section III, the chiral Schwinger
model with one-parameter class Faddeevian regularization will be
introduced following so closely the original
presentation\cite{AINW}. Furthermore, this model will be
investigated through the symplectic method\cite{JF,JW}, displaying
its chiral noninvariant nature, and the Dirac's brackets among the
phase-space coordinates will be also computed.

In section IV, the chiral Schwinger model with one-parameter
Faddeevian regularization will be reformulated as a gauge
invariant theory with the introduction of WZ field {\it via} the
new symplectic gauge-invariant formalism, presented in the
Section II. An immediate consequence produced by this process is the
generation of the infinitesimal gauge transformation that keep the
Hamiltonian invariant and the elimination of the chiral anomaly.
In this way, we have several versions for this model, which are
described by different gauge invariant Hamiltonians, but all
dynamically equivalent.

The last section is reserved to discuss the physical meaning of
our findings together with our final comments and conclusions.

\section{General formalism}

In this section, we describe the alternative embedding technique that changes the second class nature of a constrained system to the first one. This technique follows the Faddeev and Shatashivilli idea\cite{FS} and is based on a contemporary framework that handles constrained models, namely, the symplectic formalism\cite{JF,JW}.

In order to systematize the symplectic embedding formalism, we consider a general noninvariant mechanical model whose dynamics is governed by a Lagrangian ${\cal L}(a_i,\dot a_i,t)$(with $i=1,2,\dots,N$), where $a_i$ and $\dot a_i$ are the space and velocities variables, respectively. Notice that this model does not result in the loss of generality or physical content. Following the symplectic method the zeroth-iterative first-order Lagrangian is written as

\begin{equation}
\label{2000}
{\cal L}^{(0)} = A^{(0)}_\alpha\dot\xi^{(0)\alpha} - V^{(0)},
\end{equation}
where the symplectic variables are

\be
\xi^{(0)\alpha} =  \left\{ \begin{array}{ll}
                               a_i, & \mbox{with $\alpha=1,2,\dots,N $} \\
                               p_i, & \mbox{with $\alpha=N + 1,N + 2,\dots,2N .$}
                           \end{array}
                     \right.
\ee
with $A^{(0)}_\alpha$ are the one-form canonical momenta and $V^{(0)}$ is the symplectic potential. The symplectic tensor is given by

\begin{eqnarray}
\label{2010}
f^{(0)}_{\alpha\beta} = {\partial A^{(0)}_\beta\over \partial \xi^{(0)\alpha}}
-{\partial A^{(0)}_\alpha\over \partial \xi^{(0)\beta}}.
\end{eqnarray}
If this symplectic matrix is singular, it has a zero-mode $(\nu^{(0)})$ which can generate a new constraint when contracted with the gradient of symplectic potential,

\begin{equation}
\label{2020}
\Omega^{(0)} = \nu^{(0)\alpha}\frac{\partial V^{(0)}}{\partial\xi^{(0)\alpha}}.
\end{equation}
This constraint is introduced into the zeroth-iterative Lagrangian, Eq.(\ref{2000}), through a Lagrange multiplier $\eta$, generating the next one

\begin{eqnarray}
\label{2030}
{\cal L}^{(1)} &=& A^{(0)}_\alpha\dot\xi^{(0)\alpha} + \dot\eta\Omega^{(0)}- V^{(0)},\nonumber\\
&=& A^{(1)}_\gamma\dot\xi^{(1)\gamma} - V^{(1)},\end{eqnarray}
with $\gamma=1,2,\dots,(2N + 1)$ and

\begin{eqnarray}
\label{2040}
V^{(1)}&=&V^{(0)}|_{\Omega^{(0)}= 0},\nonumber\\
\xi^{(1)_\gamma} &=& (\xi^{(0)\alpha},\eta),\\
A^{(1)}_\gamma &=&(A^{(0)}_\alpha, \Omega^{(0)}).\nonumber
\end{eqnarray}
As a consequence, the first-iterative symplectic tensor is computed as

\begin{eqnarray}
\label{2050}
f^{(1)}_{\gamma\beta} = {\partial A^{(1)}_\beta\over \partial \xi^{(1)\gamma}}
-{\partial A^{(1)}_\gamma\over \partial \xi^{(1)\beta}}.
\end{eqnarray}
If this tensor is nonsingular, the iterative process stops and the Dirac's brackets among the phase space variables are obtained from the inverse matrix $(f^{(1)}_{\gamma\beta})^{-1}$ and, consequently, the Hamilton equation of motion can be computed and solved as well, as well discussed in \cite{gotay}. It is well known that a physical system can be described in terms of a symplectic manifold $M$, classically at least. From a physical point of view, $M$ is the phase space of the system while a nondegenerate closed 2-form $f$ can be identified as being the Poisson bracket. The dynamics of the system is  determined just specifying a real-valued function (Hamiltonian) $H$ on phase space, {\it i.e.}, one these real-valued function solves the Hamilton equation, namely,
\be
\label{2050a1}
\iota(X)f=dH,
\ee
and the classical dynamical trajectories of the system in phase space are obtained. It is important to mention that if $f$ is nondegenerate, Eq.(\ref{2050a1}) has an unique solution. The nondegeneracy of $f$ means that the linear map $\flat:TM\rightarrow T^*M$ defined by $\flat(X):=\flat(X)f$ is an isomorphism, due to this, the Eq.(\ref{2050a1}) is solved uniquely for any Hamiltonian $(X=\flat^{-1}(dH))$. On the contrary, the tensor has a zero-mode and a new constraint arises, indicating that the iterative process goes on until the symplectic matrix becomes nonsingular or singular. If this matrix is nonsingular, the Dirac's brackets will be determined. In Ref.\cite{gotay}, the authors consider in detail the case
when $f$ is degenerate, which usually arises when constraints are presented on the system. In which case, $(M,f)$ is called presymplectic manifold. As a consequence, the Hamilton equation, Eq.(\ref{2050a1}), may or may not possess solutions, or possess nonunique solutions. Oppositely, if this matrix is singular and the respective zero-mode does not generate a new constraint, the system has a symmetry.

The systematization of the symplectic embedding formalism begins by assuming that the gauge invariant version of the general Lagrangian $({\tilde {\cal L}}(a_i,\dot a_i,t))$ is given by
\be
\label{2051}
{\tilde{\cal  L}}(a_i,\dot a_i,\varphi_p,t) = {\cal L}(a_i,\dot a_i,t) + {\cal L}_{WZ}(a_i,\dot a_i,\varphi_p),\,\,\,\,(p = 1,2),
\ee
where $\varphi_p = (\theta, \dot\theta)$ and the extra term $({\cal L}_{WZ})$ depends on the original $(a_i,\dot a_i)$ and WZ $(\varphi_p)$ configuration variables. Indeed, this WZ Lagrangian can be expressed as an expansion in orders of the WZ variable $(\varphi_p)$ such as
\be
\label{wz1}
{\cal L}_{WZ}(a_i,\dot a_i,\varphi_p) = \sum_{n=1}^\infty \upsilon^{(n)}(a_i,\dot a_i, \varphi_p),\;\;{\text with}\;\; \upsilon^{(n)}(\varphi_p)\sim \varphi_p^{n},
\ee
which satisfies the following boundary condition,
\be
\label{wz2}
{\cal L}_{WZ} (\varphi_p=0)= 0.
\ee

The reduction of the Lagrangian, Eq.(\ref{2051}), into its first order form precedes the beginning of conversion process, thus
\be
\label{2052}
{\tilde{\cal L}}^{(0)} = A^{(0)}_{\alpha}\dot{\xi}^{(0){\alpha}} + \pi_\theta\dot\theta - {\tilde V}^{(0)},
\ee
where $\pi_\theta$ is the canonical momentum conjugated to the WZ variable, that is,
\be
\label{2052aa2}
\pi_\theta = \frac{\partial{\cal L}_{WZ}}{\partial\dot\theta} = \sum_{n=1}^\infty \frac{\partial\upsilon^{(n)}(a_i, \dot a_i,\varphi_p)}{\partial\dot\theta}.
\ee

The expanded symplectic variables are $\tilde\xi^{(0){\tilde \alpha}} \equiv (a_i, p_i,\varphi_p)$ and the new symplectic potential becomes
\be
\label{2053}
{\tilde V}^{(0)} = V^{(0)} + G(a_i,p_i,\lambda_p),\,\,\,(p = 1,2),
\ee
where $\lambda_p=(\theta,\pi_\theta)$. The arbitrary function $G(a_i,p_i,\lambda_p)$ is expressed as an expansion in terms of the WZ fields, namely
\begin{equation}
\label{2054}
G(a_i,p_i,\lambda_p)= \sum_{n=0}^\infty{\cal G}^{(n)}(a_i,p_i,\lambda_p),
\ee
with
\be
\label{2054ab}
{\cal G}^{(n)}(a_i,p_i,\lambda_p) \sim \lambda_p^n.
\ee
In this context, the zeroth one-form canonical momenta are given by

\be
{\tilde A}_{\tilde\alpha}^{(0)} = \left\{\begin{array}{lll}
                                  A_{\alpha}^{(0)}, & \mbox{with $\tilde\alpha$ =1,2,\dots,N},\\
                                  \pi_\theta, & \mbox{with ${\tilde\alpha}$= N + 1},\\
                                   0, & \mbox{with ${\tilde\alpha}$= N + 2.}
                                    \end{array}
                                  \right.
\ee
The corresponding symplectic tensor, obtained from the following general relation

\begin{equation}
{\tilde f}_{\tilde\alpha\tilde\beta}^{(0)} = \frac {\partial {\tilde A}_{\tilde\beta}^{(0)}}{\partial \tilde\xi^{(0)\tilde\alpha}} - \frac {\partial {\tilde A}_{\tilde\alpha}^{(0)}}{\partial \tilde\xi^{(0)\tilde\beta}},
\end{equation}
is
\be
\label{2076b}
{\tilde f}_{\tilde\alpha\tilde\beta}^{(0)} = \pmatrix{ { f}_{\alpha\beta}^{(0)} & 0  & 0
\cr 0 & 0 & - 1
\cr 0 & 1 & 0},
\ee
which should be a singular matrix.

The implementation of the symplectic embedding scheme consists in computing the arbitrary function $(G(a_i,p_i,\lambda_p))$. To this end, the correction terms in order of $\lambda_p$, within by ${\cal G}^{(n)}(a_i,p_i,\lambda_p)$, must be computed as well. If the symplectic matrix, Eq.(\ref{2076b}), is singular, it has a zero-mode $\tilde\varrho$ and, consequently, we have
\begin{equation}
\label{2076}
\tilde\varrho^{(0)\tilde\alpha}{\tilde f}^{(0)}_{\tilde\alpha\tilde\beta} = 0,
\end{equation}
where we assume that this zero-mode is
\begin{equation}
\label{2076a}
\tilde\varrho^{(0)}=\pmatrix{\gamma^\alpha & 0 & 0},
\end{equation}
where $\gamma^\alpha$, is a generic line matrix. Using the relation given in Eq.(\ref{2076}) together with Eq.(\ref{2076b}) and Eq.(\ref{2076a}), we get
\be
\label{2076c}
\gamma^\alpha{ f}_{\alpha\beta}^{(0)} = 0.
\ee

In this way, a zero-mode is obtained and, in agreement with the symplectic formalism, this zero-mode must be contracted with the gradient of the symplectic potential, namely,
\be
\label{2076c1}
\tilde\varrho^{(0)\tilde\alpha}\frac{\partial \tilde V^{(0)}}{\partial \tilde\xi^{(0)\tilde\alpha}} = 0.
\ee
As a consequence, a constraint arise as being
\be
\label{2076c2}
\Omega = \gamma^\alpha\[\frac{\partial V^{(0)}}{\partial \xi^{(0)\alpha}} + \frac{\partial G(a_i,p_i,\lambda_p)}{\partial \xi^{(0)\alpha}}\].
\ee
Due to this, the first-order Lagrangian is rewritten as
\be
\label{2077}
{\tilde{\cal L}}^{(1)} = A^{(0)}_{\alpha}\dot{\xi}^{(0){\alpha}} + \pi_\theta\dot\theta + \Omega \dot\eta - {\tilde V}^{(1)},
\ee
where ${\tilde V}^{(1)} = V^{(0)}$. Note that the symplectic variables are now $\tilde\xi^{(1)\tilde\alpha}\equiv (a_i,p_i,\eta,\lambda_p)$ (with $\tilde\alpha = 1,2,\dots,N+3$) and the corresponding symplectic matrix becomes
\be
\label{2078}
{\tilde f}_{\tilde\alpha\tilde\beta}^{(1)} = \pmatrix{ { f}_{\alpha\beta}^{(0)} &  {f}_{\alpha\eta} & 0 & 0
\cr {f}_{\eta\beta} & 0 & {f}_{\eta\theta} & {f}_{\eta\pi_\theta}
\cr 0 & {f}_{\theta\eta} & 0 & -1
\cr 0 & {f}_{\pi_\theta\eta} & 1 & 0},
\ee
where
\ba
\label{2078a}
{f}_{\eta\theta} &=& -\frac{\partial}{\partial\theta}\[ \gamma^\alpha\(\frac{\partial V^{(0)}}{\partial \xi^{(0)\alpha}} + \frac{\partial G(a_i,p_i,\lambda_p)}{\partial \xi^{(0)\alpha}}\)\]
,\nonumber\\
{f}_{\eta\pi_\theta} &=& -\frac{\partial}{\partial{\pi_\theta}}\[\gamma^\alpha\( \frac{\partial V^{(0)}}{\partial \xi^{(0)\alpha}} + \frac{\partial G(a_i,p_i,\lambda_p)}{\partial \xi^{(0)\alpha}}\)\],\\
{f}_{\alpha\eta} &=& \frac{\partial \Omega}{\partial \xi^{(0)\alpha}} = \frac{\partial}{\partial \xi^{(0)\alpha}}\[\gamma^\alpha\(\frac{\partial V^{(0)}}{\partial\xi^{(0)\alpha}} + \frac{\partial G(a_i,p_i,\lambda_p)}{\partial\xi^{(0)\alpha}}\)\]
.\nonumber
\ea

Since our goal is to unveil a WZ symmetry, this symplectic tensor must be singular, consequently, it has a zero-mode, namely,
\be
\label{2078b}
\tilde \nu^{(1)}_{(\nu)(a)} = \pmatrix{\mu^\alpha_{(\nu)} & 1 & a & b},
\ee
which satisfies the relation
\be
\label{2078c}
{\tilde \nu}^{(1)\tilde\alpha}_{(\nu)(a)}{\tilde f}_{\tilde\alpha\tilde\beta}^{(1)}  = 0.
\ee
Note that the parameters $(a,b)$ can be 0 or 1 and $\nu$ indicates the number of choices for ${\tilde \nu}^{(1)\tilde\alpha}$\footnote{It is important to notice that $\nu$ is not a fixed parameter}. As a consequence, there are two independent set of zero-modes, given by
\ba
\label{2078d}
\tilde \nu^{(1)}_{(\nu)(0)} &=& \pmatrix{\mu^\alpha_{(\nu)} & 1 & 0 & 1},\nonumber\\
\tilde \nu^{(1)}_{(\nu)(1)} &=& \pmatrix{\mu^\alpha_{(\nu)} & 1 & 1 & 0}.
\ea
Note that the matrix elements $\mu^\alpha_{(\nu)}$ present some arbitrariness which can be fixed in order to disclose a desired WZ gauge symmetry. In addition, in our formalism the zero-mode $\tilde\nu^{(1)\tilde\alpha}_{(\nu)(a)}$ is the gauge symmetry generator, which allows to display the symmetry from the geometrical point of view. At this point, we call attention upon the fact that this is an important characteristic since it opens up the possibility to disclose the desired hidden gauge symmetry from the noninvariant model. Different choices of the zero-mode generates different gauge invariant versions of the second class system, however, these gauge invariant descriptions are dynamically equivalent, {\it i.e.}, there is the possibility to relate this set of independent zero-modes, Eq.(\ref{2078d}), through canonical transformation $(\tilde {\bar\nu}^{(\prime,1)}_{(\nu)(a)} = T . \tilde {\bar\nu}^{(1)}_{(\nu)(a)})$ where bar means transpose matrix. For example,
\be
\pmatrix{\mu^\alpha_{(\nu)} \cr 1 \cr 0 \cr 1} = \pmatrix{1 & 0 & 0 & 0\cr 0 & 1 & 0 & 0\cr
0 & 0 & 0 & 1\cr 0 & 0 & 1 & 0} . \pmatrix{\mu^\alpha_{(\nu)} \cr 1 \cr 1 \cr 0}.
\ee
While, in the context of the BFFT formalism, different choices for the degenerated matrix $X$\cite{BN} leads to different gauge invariant version of the second class model. It is important to mentioned here that, in Ref.\cite{BN}, a suitable interpretation and explanation about this result was not given and, also, the author pointed out that not all solutions of the first step of the BFFT method can lead to a solution in the second one, which jeopardize the BFFT embedding process. From the symplectic embedding formalism, this kind of problem can be clarified and understood as well: ({\it i}) first, some choices for the degenerated matrix $X$ lead to different gauge invariant version of the second class model, however, they are dynamically equivalent, as shown by the symplectic formalism; ({\it ii}) second, some choices for the degenerated matrix $X$ can generate solutions in the first step of the BFFT method that can not lead to a pleasant solution in the second one, which hazards this WZ embedding process. This is interpreted by the symplectic point of view as been the impossibility to introduce some gauge symmetries into the model and, as consequence, an infinite numbers of WZ counter-terms in Hamiltonian\cite{BN} are required. This also happens in the iterative constraint conversion\cite{IJMP}, since there is an arbitrariness to change the second class nature of the constraints in first one. Now, it becomes clear that the arbitrariness presents on the BFFT and iterative constraint conversions methods has its origin on the choice of the zero-mode, which generates the desired WZ gauge symmetry.

From relation, Eq.(\ref{2078c}), together with Eq.(\ref{2078}) and Eq.(\ref{2078b}), some differential equations involving  $G(a_i,p_i,\lambda_p)$ are obtained, namely,
\ba
\label{2078e}
0 &=& \mu^\alpha_{(\nu)}{ f}_{\alpha\beta}^{(0)} + { f}_{\eta\beta},\nonumber\\
0 &=& \mu^\alpha_{(\nu)}{ f}_{\alpha\eta}^{(0)} + a { f}_{\theta\eta} + b { f}_{\pi_\theta\eta},\nonumber\\
0 &=& { f}_{\eta\theta}^{(0)} + b,\\
0 &=& { f}_{\eta\pi_\theta}^{(0)} - a\nonumber
\ea
Solving the relations above, some correction terms, within \break $\sum_{m=0}^\infty {\cal G}^{(m)}(a_i,p_i,\lambda_p)$, can be determined, also including the boundary conditions $({\cal G}^{(0)}(a_i, p_i,\lambda_p = 0 ))$.

In order to compute the remaining corrections terms of $G(a_i,p_i,\lambda_p)$, we impose that no more constraints arise from the contraction of the zero-mode $(\tilde\nu^{(1)\tilde\alpha}_{(\nu)(a)})$ with the gradient of potential ${\tilde V}^{(1)}(a_i,p_i,\lambda_p)$. This condition generates a general differential equation, which reads as
\begin{eqnarray}
\label{2080}
0 &=& \tilde\nu^{(1)\tilde\alpha}_{(\nu)(a)}\frac{\partial {\tilde  V}^{(1)}(a_i,p_i,\lambda_p)}{\partial{\tilde\xi}^{(1)\tilde\alpha}}
\nonumber\\
&=& \mu^\alpha_{(\nu)} \left[\frac{\partial {V}^{(1)}(a_i,p_i)}{\partial{\xi}^{(1)\alpha}} + \frac{\partial  G(a_i,p_i,\theta,\pi_\theta)}{\partial{\xi}^{(1)\alpha}}\right] + a \frac{\partial G(a_i,p_i,\lambda_p )}{\partial\theta} + b\frac{\partial G(a_i,p_i,\lambda_p)}{\partial\pi_\theta}\nonumber\\
&=& \mu^\alpha_{(\nu)} \left[\frac{\partial {V}^{(1)}(a_i,p_i)}{\partial{\xi}^{(1)\alpha}} + \sum_{m=0}^\infty\frac{\partial {{\cal G}}^{(m)}(a_i,p_i,\lambda_p )}{\partial{\xi}^{(1)\alpha}}\right] + a \sum_{n=0}^\infty\frac{\partial {\cal G}^{(n)}(a_i,p_i,\lambda_p )}{\partial\theta}\nonumber\\
&+& b \sum_{m=0}^\infty\frac{\partial {{\cal G}}^{(n)}(a_i,p_i,\lambda_p )}{\partial\pi_\theta}.
\end{eqnarray}
The last relation allows us to compute all correction terms in order of $\lambda_p$, within ${\cal G}^{(n)}(a_i,p_i,\lambda_p)$. Note that this polynomial expansion in terms of $\lambda_p$ is equal to zero, subsequently, all the coefficients for each order in this WZ variables must be identically null. In view of this, each correction term in orders of $\lambda_p$ can be determined as well. For a linear correction term, we have
\be
\label{2090}
0 = \mu^\alpha_{(\nu)} \left[\frac{\partial {V}^{(0)}(a_i,p_i)}{\partial{\xi}^{(1)\alpha}} + \frac{\partial {{\cal G}}^{(0)}(a_i,p_i)}{\partial{\xi}^{(1)\alpha}}\right] + a \frac{\partial{\cal G}^{(1)}(a_i,p_i,\lambda_p)}{\partial\theta} + b\frac{\partial{{\cal G}}^{(1)}(a_i,p_i,\lambda_p)}{\partial\pi_\theta},
\ee
where the relation $V^{(1)} = V^{(0)}$ was used. For a quadratic correction term, we get
\be
\label{2095}
0 = {\mu}^{\alpha}_{(\nu)}\left[\frac{\partial{\cal G}^{(1)}(a_i,p_i,\lambda_p)}{\partial{\xi}^{(0)\alpha}} \right] + a \frac{\partial{\cal G}^{(2)}(a_i,p_i,\lambda_p)}{\partial\theta}  + b \frac{\partial{{\cal G}}^{(2)}(a_i,p_i,\lambda_p)}{\partial\pi_\theta}.
\ee
From these equations, a recursive equation for $n\geq 2$ is proposed as

\be
\label{2100}
0 = {\mu}^{\alpha}_{(\nu)}\left[\frac{\partial {\cal G}^{(n - 1)}(a_i,p_i,\lambda_p)}{\partial{\xi}^{(0)\alpha}} \right] + a\frac{\partial{\cal G}^{(n)}(a_i,p_i,\lambda_p)}{\partial\theta} + b \frac{\partial{{\cal G}}^{(n)}(a_i,p_i,\lambda_p)}{\partial\pi_\theta},
\ee
which allows us to compute the remaining correction terms in order of $\theta$ and $\pi_\theta$. This iterative process is successively repeated up to Eq.(\ref{2080}) becomes identically null. Then, the new symplectic potential is written as
\be
\label{2110}
{\tilde  V}^{(1)}(a_i,p_i,\lambda_p) = V^{(0)}(a_i,p_i) + G(a_i,p_i,\lambda_p).
\end{equation}
Due to this, the gauge invariant Hamiltonian is obtained explicitly and the zero-mode ${\tilde\nu}^{(1)\tilde\alpha}_{(\nu)(a)}$ is identified as being the generator of the infinitesimal gauge transformation, given by

\begin{equation}
\label{2120}
\delta{\tilde\xi}^{\tilde\alpha}_{(\nu)(a)} = \varepsilon{\tilde\nu}^{(1)\tilde\alpha}_{(\nu)(a)},
\end{equation}
where $\varepsilon$ is an infinitesimal parameter.

\section{Realization of the Faddeevian regularization in the CSM}

In Ref.\cite{JR} the authors showed that the $\chi QED_{2}$ can be
quantized in a consistent an unitary way just including the
bosonization ambiguity parameter satisfying the condition $a \geq
1$ to avoid tachyonic excitations. Afterwards, Rajaraman\cite{RR}
studied the canonical structure of the model and showed that there
are two cases $a>1$ and $a=1$ belonged to distinct classes: the
$a=1$ case presenting four second class constraints belongs to an
unambiguous class containing only one representative, while the
$a>1$ case, presenting only two second class constraints,
represents a continuous one-parameter class. Due to the distinct
constraint structures, the $a>1$ class presents both massless
excitation and massive scalar excitation
($m^2=\frac{e^2a^2}{a-1}$), while in the other case there is only
massless excitation. It occurs because the chiral Schwinger model
with the familiar regularization $a>1$ has more physical degrees
of freedom than it would have were it gauge invariant. However, it
is not match with the Faddeev's case\cite{faddeev} whose the
commutator between the Gauss law constraint is non-zero. Here, the
second class nature of the set of constraints is due to the
Poisson bracket of $\pi_0$(canonical momentum conjugated to the
scalar potential $A_0$) and $G$(the Gauss law) becomes non-zero.

In \cite{PM} the author showed that the Poisson bracket involving
the Gauss law constraint is non-zero, indeed exhibits the
Faddeev's anomaly. In views of this, the author concluded that the
Faddeevian regularization not belong to the class of the usual
regularizations. In this new scenario, the gauge field is once
again a massive excitation, but the massless fermion that is
present has, unlike the usual case, a definite chirality opposite
to that entering the interaction with the electromagnetic field.
In this work, Mitra showed that with an appropriated choice of the
regularization mass term it is possible to close the second class
algebra with only three second class constraints. Although this
model is not manifestly Lorentz invariant, the Poincar\'e
generators have been constructed \cite{PM} and shown to close the
relativistic algebra on-shell. The main feature of this new
regularization is the presence of a Schwinger term in the Poisson
bracket algebra of the Gauss law, which limits the set to only
three second class constraints. To see this we start with the CSM
Lagrangian with Faddeevian regularization proposed in \cite{AINW},
reads as

\be {\cal L} = -\frac 14 \, F_{\mu\nu}\,F^{\mu\nu} +
            \frac{1}{2}\,\partial_\mu\phi\,\partial^{\mu}\phi +
            q\,\left(g^{\mu\nu} + b\,\epsilon^{\mu\nu}\right)\,
            \partial_{\mu}\phi\,A_{\nu} +
           \frac{1}{2}\,q^2\,A_{\mu} M^{\mu\nu} A_{\nu}\,,
\label{01} \ee
where the Mitra's regulator was properly
generalized. Here, $F_{\mu\nu} =
\partial_{\mu}A_{\nu}-\partial_{\nu}A_{\mu}$, $g^{\mu\nu} =
\mbox{diag}(+1,-1)$ and $\epsilon^{01} = -\epsilon^{10} =
\epsilon_{10} = 1$. $b$ is a chirality parameter, which can assume
the values $b=\pm 1$. The mass-term matrix $M_{\mu\nu}$ is defined
as

\ba M^{\mu\nu}\,=\,\pmatrix{1 & \alpha \cr
                        \alpha & \beta \cr}.
\label{02} \ea
To resemble the  Rajaraman's $a=1$ class it was chosen unity
coefficient for $A_{0}^{2}$ term. The Rajaraman's class is a
singular case in the {\it space of parameters} since its canonical
description has the maximum number of constraints with no massive
excitation. This case is reproduced in Eq.(\ref{01}) if $\alpha =
0$ and $\beta=-1$ in Eq.(\ref{02}). However, a new class appears if
we assume a nonvanishing value for $\alpha$. For example, with
Mitra's choice, $\alpha = -1$ and $\beta = -3$, the photon once
again becomes massive($m^2=4 \, q^2$), but the remaining massless
fermion has a definite chirality, opposite to that entering the
interaction with the electromagnetic field. Although this
particular choice is too restrictive, another choices are also
possible, that leads, eventually, to a new and interesting
consequences. In this work the coefficients $\alpha$ and $\beta$
are arbitrary {\it ab initio}, but the mass spectrum will impose a
constraint between them. This is verified using the symplectic
method\cite{JF,JW}.

Afterhere, the symplectic method will be used to quantize the
original second class model, obtaining the Dirac's brackets and
the respective reduced Hamiltonian as well. In order to implement
the symplectic method, the original second-order Lagrangian in the
velocity, given in Eq.(\ref{01}), is reduced into its first-order,
namely,

\be \label{03} {\cal L}^{(0)} =\pi _\phi \dot{\phi} + \pi ^1
\dot{A_1} - U^{(0)},
\end{equation}
where the symplectic potential $U^{(0)}$ is

\begin{eqnarray}
\label{04}
U^{(0)}&=& {1\over 2}(\pi _1^2 +\pi _\phi ^2 +\phi ^{\prime 2}) - A_0( \pi _1^\prime + q^2(\alpha -b)A_1 + q\pi _\phi - qb\phi^\prime )\nonumber \\
&-& A_1 \(qb\pi_\phi +{1\over 2}q^2(\beta -b^2)A_1 - q\phi^\prime
\),
\end{eqnarray}
where prime represents spatial derivative. 

The zeroth-iterative symplectic tensor is given by

\begin{equation}
\label{07} f^{(0)}(x,y)= \left( \begin{array}{ccccc}
0 & -1 & 0 & 0 & 0 \\
1 & 0 & 0 & 0 & 0 \\
0 & 0 & 0 & 0 & 0 \\
0 & 0 & 0 & -1 & 0 \\
0 & 0 & 1 & 0 & 0
\end{array} \right )\delta^2 (x - y).
\end{equation}
This matrix is obviously singular, thus, it has the following
zero-mode,

\begin{equation}
\label{08} \nu^{(0)} =\pmatrix{
0 & 0 & 1 & 0 & 0 },
\end{equation}
that when contracted with the gradient of the potential $U^{(0)}$
generates a constraint, given by,

\begin{eqnarray}
\label{09}
\Omega _1 &=&\int  \nu_\alpha ^{(0)}(x){{\partial U^{(0)}(y)}\over {\partial \xi _\alpha ^{(0)}(x)}}\,\,{\rm d} y \nonumber \\
&=&\pi _1^\prime + q^2(\alpha - b)A_1 + q\pi _\phi - qb\phi
^\prime,
\end{eqnarray}
that is identified as being the Gauss law, which satisfies the
following Poisson algebra,

\begin{equation}
\label{10} \lbrace\Omega _1(x),\Omega _1(y)\rbrace = -
2q^2\alpha\partial_x\delta^2 (x - y),
\end{equation}
where $\partial_x$ represents $\frac{\partial}{\partial x}$. The
corresponding bracket in the familiar regularization
scheme\cite{JR} is zero. That is why the Mitra's model is the one
which is in accordance with the Faddeev's scenario\cite{faddeev}
in which the Gauss law commutator has a chiral anomaly.

Bringing back the constraint $\Omega _1$ into the canonical sector
of the first-order Lagrangian by means of a Lagrange multiplier
$\eta $, we get the first-iterative Lagrangian ${\cal L} ^{(1)}$,
reads as

\begin{equation}
\label{11} {\cal L }^{(1)} =\pi _\phi \dot{\phi} + \pi ^1
\dot{A_1} + \Omega_1 \dot {\eta } - U^{(1)},
\end{equation}
with the first-order symplectic potential,

\begin{eqnarray}
\label{12} U^{(1)}={1\over 2}(\pi _1^2 + \pi _\phi ^2 + \phi
^{\prime 2}) - A_1 \(qb\pi _\phi + {1\over 2}q^2(\beta - b^2)A_1 -
q\phi^\prime\).
\end{eqnarray}

The corresponding matrix $f^{(1)}(x,y)$ is then\footnote{$\sigma
=\alpha -b$}

\begin{equation}
\label{13.1} f^{(1)}(x, y)= \left ( \begin{array}{cccccc}
0 & -1 & 0 & 0 & -qb\partial_y \\
1 & 0 & 0 & 0  & q \\
0 & 0 & 0 & -1 & q^2\sigma \\
0 & 0 & 1 & 0 & \partial_y \\
qb\partial _x & - q & - q^2\sigma & - \partial_x  & 0
\end{array} \right )\delta^2  (x-y),
\end{equation}
that is singular and has a zero-mode, given by,

\begin{equation}
\label{14} \nu^{(1)} =\pmatrix{ -q & -qb\partial_x &
-\partial_x & q^2\sigma  & 1 },
\end{equation}
that generates the following constraint,

\begin{eqnarray}
\label{15}
\Omega _2 &=& \int  \nu_\alpha ^{(1)}(x){{\partial U^{(1)}(y)}\over {\partial \xi _\alpha ^{(1)}(x)}}{\rm d} y\nonumber \\
&=& q^2\sigma\pi _1 + q^2(\beta + 1)A_1^\prime.
\end{eqnarray}
The twice-iterated Lagrangian, obtained after including the
constraint, given in Eq.(\ref{15}), into Lagrangian, Eq.(\ref{11}), by means of a
Lagrange multiplier $\zeta $, reads

\begin{equation}
\label{16} {\cal L} ^{(2)} = \pi _\phi \dot{\phi} + \pi ^1
\dot{A_1} + \Omega_1 \dot {\eta } + \Omega_2\dot {\zeta} -
U^{(2)},
\end{equation}
where $U^{(2)} = U^{(1)}\mid_{\Omega_2 = 0}$. 

The matrix $f^{(2)}(x, y)$ is\footnote{$\varrho =\beta + 1$}

\begin{equation}
\label{18} f^{(2)}(x, y)= \left ( \begin{array}{ccccccc}
0 & -1 & 0 & 0  & -qb\partial_y & 0 \\
1 & 0 & 0 & 0  & q & 0 \\
0 & 0 & 0 & -1 & q^2\sigma & q^2\varrho \partial _y \\
0 & 0 & 1 & 0 & \partial _y & q^2\sigma \\
qb\partial _x & -q & -q^2\sigma & -\partial _x  & 0 & 0 \\
0 & 0 & -q^2\varrho \partial _x &
-q^2\sigma  &  0 & 0
\end{array} \right )\delta(x-y),
\end{equation}
that is a nonsingular matrix. Then we can identify it as the
symplectic tensor of the constrained theory. The inverse of
$f^{(2)}(x, y)$ gives, after a straightforward calculation, the
Dirac brackets among the physical fields,

\begin{eqnarray}
\label{18.1} \left\{\,\phi(x)\,,\,\phi(y)\,\right\}^{*}&=&
-\frac{1}{2\,\alpha}\,
                                       \Theta (x - y)\, , \nonumber \\
\left\{\,\phi(x)\,,\,\pi_\phi(y)\,\right\}^{*}&=& \frac{(2\alpha -
b)}{2\alpha}
                                       \delta (x - y)\, , \nonumber \\
\left\{\,\phi(x)\,,\, A_1(y)\,\right\}^{*}&=& -\frac{1}{2 q
\alpha}\,
                                       \delta (x - y)\, , \nonumber \\
\left\{\,\phi(x)\,,\,\pi_1(y)\,\right\}^{*}&=&
\frac{q\sigma}{2\,\alpha}\,
                                       \Theta (x - y)\, , \nonumber \\
\left\{\pi_\phi(x),\pi_\phi(y)\right\}^{*} &=&
\frac{1}{2\,\alpha}\,
                                       \partial_x\delta (x - y)\, , \nonumber \\
\left\{\pi_\phi(x),A_1(y)\right\}^{*} &=& \frac{b}{2 q\,\alpha}
                                       \partial_x \delta (x - y)\, , \nonumber \\
\left\{\pi_\phi(x),\pi_1(y)\right\}^{*} &=& -
\frac{qb\sigma}{2\alpha}
                                       \delta (x - y)\, , \nonumber \\
\left\{A_{1}(x),A_{1}(y)\right\}^{*} &=&
\frac{1}{2\,q^{2}\,\alpha}\,
                                        \partial_x\delta (x - y)\, , \\
\left\{A_{1}(x),\pi_{1}(y)\right\}^{*} &=& \left(\frac{\alpha +
b}{2\,\alpha}
                                           \right)\delta (x - y)\, ,\nonumber \\
\left\{\pi_{1}(x),\pi_{1}(y)\right\}^{*} &=&
-\frac{q^{2}\sigma^2}{2\,\alpha}
                                           \Theta (x - y)\,\,, \nonumber
\end{eqnarray}
where $\Theta(x - y)$ represents the sign function. This means
that the Mitra model is not a gauge invariant theory. The
second-iterative symplectic potential $U^{(2)}$ is identified as
the reduced Hamiltonian, given by,
\begin{eqnarray}
\label{18.2} H_{r} = \int &dx& \left\{ \frac{1}{2}\pi_{1}^{2} +
\alpha\,\pi_{1}A_{1}^{\prime} + q\left(1 -
b\,\alpha\right)A_1\phi^{\prime} + \phi^{'2} +\right.\nonumber\\
&+&\left.\frac{b}{q}\,\phi^{\prime}\pi_{1}^{\prime}  +
\frac{1}{2\,q^{2}} \pi_{1}^{\prime 2}
+\frac{1}{2}\,q^{2}\left(\alpha^{2}-
\beta\right)A_{1}^{2}\right\}\, ,
\end{eqnarray}
where the constraints $\Omega_1$ and $\Omega_2$ were assumed equal
to zero in a strong way.

At this stage, we are interested to compute the energy spectrum.
To this end, we use the reduced Hamiltonian, Eq.(\ref{18.2}), and the
Dirac's brackets, Eq.(\ref{18.1}), to obtain the following equations of
motion for the fields,

\begin{eqnarray}
\label{18.3} \dot{\phi} &=& b\,\phi^{\prime} - \frac{1}{q}
\pi_{1}^{\prime} + \frac{q}{2\,\alpha}
\left(1 - 2\alpha^{2} + \beta\right)A_{1}, \nonumber \\
\dot{\pi}_{1} &=& -b \pi_{1}^{\prime}
+\frac{q^{2}}{2\,\alpha}\left[(b-\alpha)
(1-\alpha^{2}) - (b + \alpha)(\alpha^{2} - \beta)\right]A_{1},\\
\dot{A}_{1} &=& \left(\frac{\alpha + b}{2\alpha}\right)\pi_{1} -
\left(\frac{1 + \beta}{2\alpha}\right)A_{1}^{\prime}\,\, .
\nonumber
\end{eqnarray}

Now, we are ready to determine the spectrum of the model.
Isolating $\pi^{1}$ from the constraint $\Omega_{2}$, given in
Eq.(\ref{15}), and substituting in Eq.(\ref{18.3}), we
get

\begin{eqnarray}
\label{18.4} \left(\frac{2\,\alpha}{\alpha + b}\right)\ddot{A}_{1}
&+& b\left(\frac{1 + \beta}{\alpha +
b}\right)A_{1}^{\prime\prime} = -\left(\frac{2\,b\,\alpha}
{\alpha + b} + \frac{1 + \beta}{\alpha +
b}\right)
\dot{A}_{1}^{\prime}\,+\, \nonumber\\
\! & + & \!\frac{q^2}{2\alpha}\left[\left(b - \alpha\right)
\left(1 - \alpha^{2}\right) - \left(b + \alpha\right)
\left(\alpha^{2}-\beta\right)\right]A_{1}\, .
\end{eqnarray}
To get a massive Klein-Gordon equation for the photon field we
must set

\be \label{18.5} \left(1 + \beta\right) + b\,\left(2\alpha\right)
= 0\, , \ee
which relates $\alpha$ and $\beta$ and shows that the
regularization ambiguity adopted in \cite{PM} can be extended to a
continuous one-parameter class (for a chosen chirality). We have,
using Eq.(\ref{18.4}) and Eq.(\ref{18.5}), the following mass formula
for the massive excitation of the spectrum,

\be \label{18.6} m^{2} = q^{2}\,\frac{1 + b\,\alpha}{b\,\alpha}\,
. \ee
Note that to avoid tachyonic excitations, $\alpha$ is
further restricted to satisfy $b\,\alpha = |\alpha|$, so $\alpha
\rightarrow -\alpha$ interchanges from one chirality to another.
Observe that in the limit $\alpha \rightarrow 0$ the massive
excitation becomes infinitely heavy and decouples from the
spectrum. This leads us back to the four-constraints class. It is
interesting to see that the redefinition of the parameter as $a=1+
|\alpha|$ leads to,

\be \label{18.7} m^2 = \frac{q^2 a^2}{a-1}, \ee
which is the celebrate mass formula of the chiral Schwinger model,
showing that the parameter dependence of the mass spectrum is
identical to both the Jackiw-Rajaraman and the Faddeevian
regularizations.

Let us next discuss the massless sector of the spectrum. To
disclose the presence of the chiral excitation we need to
diagonalize the reduced Hamiltonian, Eq.(\ref{18.2}). This procedure
may, at least in principle, impose further restrictions over
$\alpha$. This all boils down to find the correct  linear
combination of the fields leading to the free chiral equation of
motion. To this end we substitute $\pi^{1}$ from its definition
and $A_{1}$ from the Klein-Gordon equation into
Eq.(\ref{18.3}) to obtain

\begin{eqnarray}
\label{18.8} 0 &=&\frac{\partial}{\partial t}\left\{ \phi  +
\frac{q}{2\alpha} \left(\frac{2 + 2\,b\,\alpha -
\alpha^{2}}{m^{2}}\right) \dot{A}_{1} +
\frac{1}{q}\left(\frac{\alpha}{\alpha + b}
\right)A_{1}^{\prime}\right\}\,-\,  \\
&-& \frac{\partial}{\partial x}\left\{ b\,\phi -
\frac{1}{q}\left(\frac{\alpha}{\alpha + b}\right)\dot{A}_{1} +
\left[\frac{q}{2\alpha}\left(\frac{2 + 2\,b\,\alpha -
\alpha^{2}}{m^{2}}\right)
-\frac{1}{q}\left(\frac{2\,b\,\alpha}{\alpha +
b}\right)\right]\right\}\,\,.\nonumber
\end{eqnarray}
This expression becomes the equation of motion for a self-dual
boson $\chi$, given by,

\be \dot{\chi} - b\,\chi^{\prime} = 0, \label{18.9} \ee
if we identify the coefficients for $\dot A_1$ and $A_1^{\prime}$
in the two independent terms of Eq.(\ref{18.8}) with,

\be \label{18.10} \chi = \phi  +  \frac 1{q}
\left(\frac{\alpha}{\alpha + b}\right)\left(A_{1}^{\prime}- b
\dot{A}_{1}\right). \ee
This field redefinition, differently from the case of the massive
field whose construction imposed condition, Eq.(\ref{18.5}), does not
restrain the parameter $\alpha$ any further. Using the constraints
$\Omega_1$ and $\Omega_2$, given in Eq.(\ref{09}) and Eq.(\ref{15})
respectively, and Eq.(\ref{18.9}), all the fields can be expressed
as functions of the free massive scalar $A_1$ and the free chiral
boson $\chi$, interpreted as the bosonized Weyl fermion. The main
result of this section is now complete, i.e., the construction of
the one-parameter class regularization generalizing Mitra's
proposal. In the next section, this general model will be
reformulate as a gauge theory in order to eliminate the chiral
anomaly.

\section{Embedding the chiral Schwinger model with the Faddeevian regularization}

In order to begin with the WZ embedding formulation, some WZ
counter-terms, embraced by ${\cal L}_{WZ}$, are introduced into
the original Lagrangian, leading to the gauge invariant
Lagrangian, namely,

\be \label{28} \widetilde{\cal L} ={\cal L} +{\cal L}_{WZ} \ee

In agreement with the symplectic embedding formalism, the
invariant Lagrangian above, Eq.(\ref{28}), must be reduced into its
first-order form, given by

\begin{equation}
\label{29} {\tilde {\cal L}}^{(0)} =\pi _\phi \dot{\phi} + \pi ^1
\dot{A_1} + \pi_\theta \dot\theta - {\tilde U}^{(0)},
\end{equation}
where

\ba \label{30} {\tilde U}^{(0)}&=& {1\over 2}(\pi _1^2 + \pi _\phi
^2 + \phi ^{\prime 2}) -  A_0 (\pi _1^\prime + q^2(\alpha - b)A_1 + q\pi _\phi - qb\phi
^\prime) \nonumber \\
&-& A_1 \(qb\pi _\phi +
{1\over 2}q^2(\beta - b^2)A_1 - q\phi^\prime \) +
G(\phi , \pi _\phi, A_0 ,A_1, \pi _1, \lambda_p), \ea where the
arbitrary function $G$ is \be \label{31} G(\phi , \pi _\phi, A_0
,A_1, \pi _1, \lambda_p)=\sum_{n=0}^{\infty} {\cal G}^{(n)}(\phi ,
\pi _\phi, A_0 ,A_1, \pi _1, \lambda_p), \ee
where the function expanded in terms the WZ variables $(\lambda_p
=(\theta,\pi_\theta ))$ is given by
\be \label{32} {\cal G}^{(n)} \sim (\lambda_p )^n. \ee

\noindent The symplectic variables are given by ${\tilde \xi} _{\tilde \alpha}^{(0)}=(
\phi ,\pi _\phi ,A_0 ,A_1 ,\pi _1, \lambda_p ).$ 
The corresponding symplectic matrix ${\tilde f}^{(0)}(x,y)$ reads

\begin{equation}
\label{34} {\tilde f}^{(0)}(x,y) = \pmatrix{0 & -1 & 0 & 0 & 0 & 0
& 0 \cr 1 & 0 & 0 & 0 & 0 & 0 & 0 \cr 0 & 0& 0& 0& 0 & 0 & 0 \cr 0
& 0 & 0 & 0 & -1 & 0 & 0 \cr 0 & 0 & 0 & 1 & 0 & 0 & 0 \cr 0 & 0 &
0 & 0 & 0 & 0 & -1\cr 0 & 0 & 0 & 0 & 0 & 1 & 0} \delta^{(2)}  (x-y).
\end{equation}
As the matrix is singular, it has a zero-mode,

\be \label{35} \widetilde{\nu}^{(0)} =\pmatrix{0 & 0 & 1 & 0 & 0 &
0 & 0 & 0} \ee which generates the following constraint

\be \label{36} \Omega =-\pi_1^{\prime} -q^2(\alpha -b)A_1
-q\pi_\phi +qb\phi^{\prime} +\int {\rm d}\omega\sum_{n=0}^{\infty}
{{\delta{\cal G}^{(n)}(\omega)}\over{\delta A_0(y)}}, \ee after
the contraction with the gradient of the symplectic potential.

Following the symplectic embedding formalism, this constraint is
introduced into the kinetical sector of the zeroth-iterative
Lagrangian, Eq.(\ref{29}), through a Lagrange multiplier $(\eta)$,
which leads to the first-iterative Lagrangian, given by

\be \label{37}
\widetilde{\cal L}^{(1)} =\pi_\phi \dot \phi +
\pi_1 {\dot A}^1 + \pi_\theta \dot\theta +\Omega\dot\eta
-\widetilde{U}^{(1)}, \ee where
$\widetilde{U}^{(1)}=\widetilde{U}^{(0)}\mid_{\Omega=0}$. Now, the
symplectic variables are ${\tilde \xi} _{\tilde \alpha}^{(1)}=( \phi ,\pi
_\phi ,A_0 ,A_1 ,\pi _1, \eta, \lambda_p )$ with the respective
symplectic matrix,

\begin{equation}
\label{38} {\tilde f}^{(1)}= \pmatrix{0 & -\delta^{(2)} (x-y) & 0 & 0
& 0 & f_{\phi\eta}^{(1)}& 0 & 0 \cr \delta^{(2)} (x-y) & 0 & 0 & 0 & 0
& f_{\pi_\phi \eta}^{(1)} & 0 & 0 \cr 0 & 0& 0& 0& 0 & f_{A_0
\eta}^{(1)} & 0 & 0 \cr 0 & 0 & 0 & 0 & -\delta^{(2)} (x-y) & f_{A_1
\eta}^{(1)} & 0 & 0 \cr 0 & 0 & 0 & \delta^{(2)} (x-y) & 0 & f_{\pi_1
\eta}^{(1)} & 0 & 0 \cr f_{\eta\phi}^{(1)}& f_{\eta\pi_\phi}^{(1)}
& f_{\eta A_0}^{(1)} & f_{\eta A_1 }^{(1)} & f_{\eta \pi_1
}^{(1)}& 0 & f_{\eta \theta}^{(1)} & f_{\eta \pi_\theta}^{(1)}\cr
0 & 0 & 0 & 0 & 0 & f_{\theta\eta}^{(1)}& 0 & -\delta^{(2)} (x-y)\cr 0
& 0 & 0 & 0 & 0 &f_{\pi_\theta \eta}^{(1)}& \delta^{(2)} (x-y) & 0}.
\end{equation}
with

\ba\label{39} f_{\phi\eta}^{(1)}&=& qb\partial_y \delta^{(2)} (x-y)
+\frac{\delta}{\delta \phi(x)}\int {\rm d}\omega
\sum_{n=0}^{\infty}\frac{\delta{\cal
G}^{(n)}(\omega)}{\delta A_0 (y)},\nonumber \\
f_{\pi_\phi \eta}^{(1)}&=&-q\delta^{(2)} (x-y) +\frac{\delta}{\delta
\pi_\phi(x)}\int {\rm d}\omega
\sum_{n=0}^{\infty}\frac{\delta{\cal
G}^{(n)}(\omega)}{\delta A_0 (y)},\nonumber \\
f_{A_0 \eta}^{(1)}&=&\frac{\delta}{\delta A_0 (x)}\int {\rm
d}\omega \sum_{n=0}^{\infty}\frac{\delta{\cal
G}^{(n)}(\omega)}{\delta A_0 (y)},\nonumber \\
f_{A_1 \eta}^{(1)}&=&-q^2 (\alpha -b)\delta^{(2)} (x-y)
+\frac{\delta}{\delta A_1 (x)}\int {\rm d}\omega
\sum_{n=0}^{\infty}\frac{\delta{\cal G}^{(n)}(\omega)}{\delta
A_0(y)},\\
f_{\pi_1 \eta}^{(1)}&=&-\partial_y \delta^{(2)} (x-y)
+\frac{\delta}{\delta \pi_1 (x)}\int {\rm d}\omega
\sum_{n=0}^{\infty}\frac{\delta{\cal
G}^{(n)}(\omega)}{\delta A_0 (y)},\nonumber \\
f_{\theta\eta}^{(1)}&=&\frac{\delta}{\delta\theta(x)}\int {\rm
d}\omega \sum_{n=0}^{\infty}\frac{\delta{\cal
G}^{(n)}(\omega)}{\delta A_0 (y)},\nonumber \\
f_{\pi_\theta \eta}^{(1)}&=&\frac{\delta}{\delta\pi_\theta
(x)}\int {\rm d}\omega \sum_{n=0}^{\infty}\frac{\delta{\cal
G}^{(n)}(\omega)}{\delta A_0 (y)},\nonumber \ea

Now, in order to unveil the gauge symmetry and to put our result in perspective with others, we make a ``educated guess" for the zero-mode, namely,

\be \label{43} \widetilde{\nu}^{(1)\widetilde{\alpha}} =
\pmatrix{q & qb\partial_x & 0 & \partial_x & 0 & 1 & \Delta &
-q^2 c \partial_x /\Delta},
\ee
where $\Delta^2 = q^2(\beta +1)$ and $c$ is a chiral parameter in the WZ sector.The contraction of this zero-mode with the symplectic
tensor leads to eight differential equations, which are
written as

\ba \label{44} 0&=&\int {\rm d}
x\frac{\delta}{\delta\phi(y)}\int_\omega \sum_{n=0}^{\infty}
\frac{\delta{\cal G}^{(n)}(\omega)}{\delta A_0(x)},\nonumber\\
0&=&\int {\rm d} x\frac{\delta}{\delta\pi_\phi(y)}\int_\omega
\sum_{n=0}^{\infty}
\frac{\delta{\cal G}^{(n)}(\omega)}{\delta A_0(x)},\nonumber\\
0&=&\int {\rm d} x\frac{\delta}{\delta A_0 (y)}\int_\omega
\sum_{n=0}^{\infty}
\frac{\delta{\cal G}^{(n)}(\omega)}{\delta A_0(x)},\nonumber\\
0&=&\int {\rm d} x \[ q^2(\alpha -b)\delta^{(2)} (x - y) -
\frac{\delta}{\delta A_1 (y)}\int_\omega \sum_{n=0}^{\infty}
\frac{\delta{\cal G}^{(n)}(\omega)}{\delta A_0(x)}\],\nonumber\\
0&=&\int {\rm d} x\frac{\delta}{\delta \pi_1 (y)}\int_\omega
\sum_{n=0}^{\infty}
\frac{\delta{\cal G}^{(n)}(\omega)}{\delta A_0(x)},\\
0&=&\int {\rm d} x \left[ -q^2(\alpha +b)\partial_x \delta^{(2)} (x -
y) +
\partial_x \frac{\delta}{\delta A_1 (y)}\int_\omega \sum_{n=0}^{\infty}
\frac{\delta{\cal G}^{(n)}(\omega)}{\delta A_0(x)} + \right. \nonumber\\
&+& \left. \Delta \frac{\delta}{\delta \theta(y)}\int_\omega
\sum_{n=0}^{\infty} \frac{\delta{\cal G}^{(n)}(\omega)}{\delta
A_0(x)} - \frac{q^2}{\Delta} c \partial_x
\frac{\delta}{\delta\pi_\theta(y)}\int_\omega \sum_{n=0}^{\infty}
\frac{\delta{\cal G}^{(n)}(\omega)}{\delta A_0(x)}
\right],\nonumber\\
0&=&\int {\rm d} x\[ -\frac{q^2}{\Delta} c \partial_x \delta^{(2)} ( x-y)
-\frac{\delta}{\delta \theta(y)}\int_\omega \sum_{n=0}^{\infty}
\frac{\delta{\cal G}^{(n)}(\omega)}{\delta
A_0(x)} \], \nonumber\\
0&=&\int {\rm d} x\[ -\Delta \delta^{(2)} ( x-y) -\frac{\delta}{\delta
\pi_\theta(y)}\int_\omega \sum_{n=0}^{\infty} \frac{\delta{\cal
G}^{(n)}(\omega)}{\delta A_0(x)} \]. \nonumber \ea

From the fourth relation above, we get the boundary condition,
which is written as

\be \label{45} {\cal G}^{(0)} = q^2(\alpha -b) A_0 A_1 .\ee
So, the zeroth correction term is
\be \label{46} {\cal G}^{(0)} = q^2(\alpha -b) A_0 A_1 + {\cal
G}^{(0)}(A_1), \ee
since from the third relation above, we see that ${\cal G}^{(0)}$ has no quadratic dependence in terms of $A_0$.
While from the seventh and eighth relation above, the first
correction term, at least, is obtained partially as being
\be \label{47} {\cal G}^{(1)} = -\frac{q^2}{\Delta} c \partial_1
\theta A_0 - \Delta\pi_\theta A_0 .\ee

It is important to notice that the relation above can not envelop
all of the first-correction terms, because some of them can not
depend on the temporal component of the potential field. In view
of this, we rewrite this term as

\be \label{48} {\cal G}^{(1)} = -\frac{q^2}{\Delta} c \partial_1
\theta A_0 - \Delta\pi_\theta A_0 +{\cal G}^{(1)}(A_1 ,\lambda_p )
.\ee
Thus, the constraint, given in Eq. (\ref{36}), becomes

\be \label{36a} 
\Omega =-\pi_1^{\prime} 
-q\pi_\phi +qb\phi^{\prime} - \frac{q^2}{\Delta} c \theta^{\prime} 
- \Delta \pi_\theta. \ee
This constraint satisfies the following Poisson algebra

\be
\label{36b}
\{\Omega(x),\;\Omega(y)\} = 2q^2(b -c) \partial_y \delta(x-y).
\ee
Note that the elimination of the chiral anomaly above imposes a condition on the chiral parameters present on the original model and on the WZ sector, $b$ and $c$, respectively, {\it i.e}, $b=c$. Hence, there is a specific set of WZ gauge symmetries that can deal with the chiral anomaly.

The symplectic potential can be expressed as
\ba \label{49} {\tilde U}^{(1)}&=&{\tilde U}^{(0)}\mid_{\Omega
=0}\nonumber\\
&=& {1\over 2}(\pi _1^2 + \pi _\phi ^2 + \phi ^{\prime 2}) - A_1
\(qb\pi _\phi + {1\over 2}q^2(\beta - b^2)A_1 - q\phi^\prime
\)\nonumber\\ &+& \sum_{n=1}^{\infty}{\cal
G}^{(n)}(A_1,\lambda_p)+{\cal G}^{(0)}(A_1). \ea

Now, it becomes necessary to guarantee that no more constraint
arises. To this end, we impose that the contraction of the
zero-mode, Eq.(\ref{43}), with the gradient of the symplectic
potential, Eq.(\ref{49}), does not produce a new constraint,
namely,
\ba\label{50}0 &=& \int \widetilde{\nu}^{(1)\widetilde{\alpha}}{{\partial
U^{(1)}(y)}\over {\partial {\tilde \xi} ^{(1)\widetilde{\alpha}}(x)}}\,\,{\rm d} y
,\nonumber\\
&=&\int {\rm d} y\[ -q^2 (\beta +1) A_1 (y) \partial_y \delta^{(2)}
(x-y) +\sum_{n=1}^{\infty} \partial_x \frac{\delta {\cal
G}^{(n)}(A_1 ,\lambda_p )(y)}{\delta A_1 (x)}+
\right. \nonumber\\
&+&\left. \partial_x \frac{\delta {\cal G}^{(0)}(A_1)(y)}{\delta
A_1 (x)} + \Delta \sum_{n=1}^{\infty}\frac{\delta {\cal
G}^{(n)}(A_1 ,\lambda_p )(y)}{\delta \theta(x)}
-\frac{q^2}{\Delta} b \partial_x \sum_{n=1}^{\infty}\frac{\delta
{\cal G}^{(n)}(A_1 ,\lambda_p )(y)}{\delta \pi_\theta (x)} \]. \ea
Which is a polynomial expression in order of the WZ fields
$(\lambda_p)$. For the zeroth relation written in terms of WZ fields, we get
\ba \label{51} 0 &=& \int {\rm d} y\[ -q^2 (\beta +1) A_1 (y)
\partial_y \delta^{(2)} (x-y) + \partial_x \frac{\delta {\cal G}^{(0)}(A_1)(y)}{\delta
A_1 (x)} +\right.\nonumber \\ &+&\left. \Delta \frac{\delta {\cal
G}^{(1)}(A_1 ,\lambda_p )(y)}{\delta \theta(x)}
-\frac{q^2}{\Delta} b \partial_x \frac{\delta {\cal G}^{(1)}(A_1
,\lambda_p )(y)}{\delta \pi_\theta (x)} \] .\ea
At this point, it is important to notice that some degeneracy
appears, since we can not solve the relation above leading to an
unique result, {\it i.e}, this relation has solution but it is not
unique. At first, this sound quite bad, however, this shows how
powerful the symplectic embedding formalism can be: different choices for ${\cal G}^{(0)}(A_1)$ and ${\cal
G}^{(1)}(A_1 ,\lambda_p )$ leads to distinct gauge invariant Hamiltonian descriptions for the
noninvariant model, but with the same WZ gauge symmetry. It is a new feature in the WZ embedding concept that could be revealed in the symplectic embedding formalism. On the other hand, this relation can lead to
a hard computation of the gauge invariant symplectic potential
just assuming a bad solution for ${\cal G}^{(0)}(A_1)$ and ${\cal
G}^{(1)}(A_1 ,\lambda_p )$.

As we are interested in comparing our result with others, we
tackle a fine solution which becomes the computation of correction
terms in WZ fields an easy task. The chosen solutions for
Eq.(\ref{51}) are
\ba \label{52} {\cal G}^{(0)}(A_1) &=&
-\frac{q^4}{2\Delta^2} A_1^2,\nonumber\\
{\cal G}^{(1)}(A_1 ,\lambda_p )&=& \Delta A_1
\partial_1 \theta - \frac{q^2}{\Delta} b A_1 \pi_\theta .
\ea
So, the symplectic potential becomes
\ba \label{53} {\tilde U}^{(1)} &=& {1\over 2}(\pi _1^2 + \pi
_\phi ^2 + \phi ^{\prime 2}) - A_1 \( qb\pi _\phi + {1\over
2}q^2(\beta - b^2)A_1 - q\phi^\prime +\right. \nonumber\\&+&
\left. \frac{q^4}{2\Delta^2} b A_1 - \Delta \partial_1 \theta 
+\frac{q^2}{\Delta} b \pi_\theta \) + \sum_{n=2}^{\infty}{\cal
G}^{(n)}(A_1,\lambda_p). \ea
Again, we use the Eq.(\ref{50}), which allows us to compute the
quadratic correction term in WZ fields,
\ba\label{54} 0 &=& \int {\rm d} y\[ \Delta \partial_1 \theta(y)
\partial_y \delta^{(2)} (x-y) -\frac{q^2}{\Delta}\pi_\theta (y)\partial_x \delta^2 ( x-y)+ \right.
\nonumber\\ &+&\left. \Delta \frac{\delta {\cal G}^{(2)}(A_1
,\lambda_p )(y)}{\delta \theta(x)} -\frac{q^2}{\Delta}\partial_x
\frac{\delta {\cal G}^{(2)}(A_1 ,\lambda_p )(y)}{\delta \pi_\theta
(x)}
\], \ea
which leads to the following solution,
\be \label{55} {\cal G}^{(2)} =\frac{q^2}{2} (\partial_1
\theta)^2 - {1\over 2}(\pi_\theta)^2 .\ee

As this last correction term ${\cal G}^{(2)}$ has dependence only
on the WZ fields, the correction terms ${\cal G}^{(n)}=0$ for
$n\geq 3$. Due to this, the symplectic potential, identified as being
the gauge invariant Hamiltonian, is
\ba \label{56} {\tilde {\cal H}} = {\tilde U}^{(1)} &=& {1\over 2}\(\pi _1^2 + \pi
_\phi ^2 +(\partial_1 \phi)^2 +(\partial_1 \theta)^2
-(\pi_\theta)^2\) \nonumber \\
&+& A_0\(-\partial_1 \pi_1 -q\pi_\phi +qb\partial_1 \phi -
\frac{q^2}{\Delta} b \partial_1 \theta -\Delta \pi_\theta \)\nonumber\\
&+& A_1 \( -qb\pi _\phi + {q^2\over 2}\frac{\beta^2}{(\beta
+1)}A_1 + q\partial_1 \phi -\frac{q^2}{\Delta} b \pi_\theta
+\Delta\partial_1\theta\), \ea
which is the same result obtained in \cite{IJMP}, when $\beta
=-a$.

To complete the gauge invariant reformulation of the model, the
infinitesimal gauge transformation will be also computed. In
agreement with the symplectic method, the zero-mode,
${\widetilde{\nu}^{(1)}}$, Eq.(\ref{43}), is the generator of the
infinitesimal gauge transformation $(\delta{\cal
O}=\varepsilon\widetilde{\nu})$, given by
\ba \label{57} \delta\phi &=&q\varepsilon, \nonumber\\
\delta\pi_\phi &=&-qb\partial\varepsilon, \nonumber\\
\delta A_0 &=&0, \nonumber\\
\delta A_1 &=&-\partial\varepsilon, \nonumber\\
\delta\pi_1 &=&0, \\
\delta\eta &=&\varepsilon, \nonumber\\
\delta\theta &=&\Delta\varepsilon, \nonumber\\
\delta\pi_\theta &=&\frac{q^2}{\Delta} b \partial\varepsilon,
\nonumber \ea
where $ \varepsilon$ is an infinitesimal time dependent
parameter.
\section{Final Discussions}

In this work the bosonized form of the CSM fermionic determinant
parameterized by a single real number, which extends early
regularizations \cite{JR,PM}, was studied from the symplectic
point of view. Afterwards, we reproduce the spectrum that has been shown
in Ref.\cite{AINW} to consist
of a chiral boson and a massive photon field. The mass formula for
the scalar excitation was shown to reproduce the Jackiw-Rajaraman
result, while the massless excitation was shown to reproduce the
Mitra result. In views of this, unitarity condition restraints the
range of the regularization parameter in a similar way. This noninvariant
model was reformulated as a gauge invariant
model {\it via} the symplectic embedding formalism, where we
were able to produce the gauge invariant version of
the model, in order to eliminate the chiral anomaly in the Gauss
law commutator.  It is important to mention here that the gauge invariant
version of CSM, given in Ref.\cite{IJMP}, can be also obtained when $\beta = - a.$
Notice that the invariant theory was obtained with
the introduction of a finite numbers of WZ terms.
It is also important to emphasize that we can obtain different
Hamiltonian formulations for the model.
Different choices of the zero-mode generates different gauge
invariant versions of the second class system, however, these
gauge invariant descriptions are dynamically equivalent, {\it i.e.},
there is the possibility to relate this set of independent zero-modes
through canonical transformation \cite{JPA}. Another interesting point
discussed in this work is the geometrical interpretation given to the degeneracy
of the matrix $X$ presents on the BFFT formalism and the arbitrariness on the iterative method.  Different choices for this matrix $X$, or way to turn second class constraints to first one in the iterative method, leads to distinct gauge invariant version of the second class model. But, from the symplectic embedding point of view, these invariant descriptions are equivalent. Besides, it was also possible to reveal an interesting feature in the symplectic embedding formulation: the possibility to introduce boundary conditions that has no dependence on the WZ variables, which opens new {\it modus} to run the embedding process. This is a new concept which lifts the WZ embedding idea to the next level.

\section{ Acknowledgments}
This work is supported in part by FAPEMIG and CNPq, Brazilian Research Agencies. One of us, W. Oliveira, would like to thank Prof. Dr. Jos\'e Maria Filardo Bassalo, his friend and teacher, for the great stimulation gave in the beginning of his career.

\end{document}